# Deterministic tuning of slow-light in photonic-crystal waveguides through the C and L bands by atomic layer deposition

Charlton J. Chen[1], Chad A. Husko[1], Inanc Meric[2], Ken L. Shepard[2], and Chee Wei Wong[1,*]

William M. J. Green[3], Yurii A. Vlasov[3], and Solomon Assefa[3,*]

[1]Optical Nanostructures Laboratory, Columbia University, New York, NY 10027

[2]Department of Electrical Engineering, Columbia University, New York, NY 10027

[3]IBM T. J. Watson Research Center, Yorktown Heights, NY 10598

**Abstract:** We demonstrate digital tuning of the slow-light regime in silicon photonic-crystal waveguides by performing atomic layer deposition of hafnium oxide. The high group-index regime was deterministically controlled (red-shift of 140 ± 10 pm per atomic layer) without affecting the group-velocity dispersion and third-order dispersion. Additionally, differential tuning of 110 ± 30 pm per monolayer of the slow-light TE-like and TM-like modes was observed. This passive post-fabrication process has potential applications including the tuning of chip-scale optical interconnects, as well as Raman and parametric amplification.

* Electronic mail: cww2104@columbia.edu; sassefa@us.ibm.com

Dramatic reduction of the group-velocity of light has been demonstrated in atomic and solid-state systems [1, 2] with greatly increased light-matter interaction, although typically at the expense of bandwidth. Slow-light in photonic-crystal (PhC) waveguides, through strong structural dispersion, allows larger bandwidth for potential applications such as optical buffering and switching [3,4,5], disordered localization [6,7], and nonclassical optics [8]. The dispersion-to-loss ratio [9] is comparable to that of single-mode optical fibers, allowing strong enhancement of nonlinear processes on the chip, such as third-harmonic generation, self-phase modulation, Raman and parametric processes [10, 11]. However, in order to operate at a particular frequency, these devices often possess stringent fabrication requirements that are difficult to achieve using current e-beam or deep-UV lithography; even slight fabrication deviations at the nanometer level can shift the tight operating bandwidths of the integrated photonic devices [12]. Active approaches to tune photonic elements include aligned external pump laser beams [13], integrated piezoelectric elements [14] or micro-heaters [3]. To maintain the shifted dispersion or resonances in active tuning approaches, a finite external power must be continuously applied to the photonic elements. Alternatively, passive tuning approaches have been examined, such as GaAs wet-etching [15], nitrogen or Xe condensation in cryostats [16], self-assembled polypeptide monolayers [17] and electron beam induced compaction [18]. Recently we examined an ALD approach to tune PhC microcavity resonances with a precision of ~ 122 pm per hafnium oxide ($HfO_2$) layer [19]; additionally, the tuning of photonic bands in silicon-on-insulator slab PhC structures has also been examined with titanium dioxide deposition [20]. Here we propose and demonstrate for the first time a passive post-fabrication scheme for tuning dispersion in slow-light PhC waveguides by utilizing a digital self-limiting deposition of $HfO_2$ monolayers.

To study the effect of passive tuning of the slow-light regime, we designed PhC waveguides and Mach-Zehnder interferometer (MZI) devices for transmission measurements in the near-infrared [3, 21]. Each MZI device consists of a Y-splitter connected to a strip waveguide on one arm and to a PhC waveguide on the other arm, as shown in Figure 1(b). The interference fringes from MZI spectral measurements are used to determine group-velocity using a procedure described below. The PhC waveguides are W0.9 line defects created by removing a single row of air holes in a hexagonal lattice of air holes along the Γ-K direction and then decreasing the defect width by 10%. The lattice parameter of the PhCs (*a*) is 410 nm with air hole radii of 108 nm (*r/a* ratio of 0.265). The structures were fabricated by e-beam lithography on SOI substrates with a silicon slab thickness of 220 nm (*t/a* ratio of 0.537). The PhC waveguide is 250 μm long and butt-coupled to strip waveguides at both ends, and has previously demonstrated low-loss of 2.4-dB/mm [22]. The underlying oxide was subsequently removed by HF etching. The silicon strip waveguides are tapered adiabatically as they connect to polymer couplers which are used for low-loss coupling to off-chip polarization-maintaining tapered-lensed fibers [22].

Figure 2(a) shows the projected band structure of our PhC waveguide, computed through 3D plane wave expansion [23]. In order to get the best fitting to experimental data a procedure described in reference 21 was used. Within the band gap, there are two TE-like modes (even and odd modes). The even mode exhibits slow-light characteristics near the band-edge where *dω/dk* becomes increasingly small, resulting in large group indices, $n_g = c\ (dk/d\omega)$. The corresponding projected band structure for the TM-like mode is shown Figure 2(b) where, although no band gap exists, one observes a Bragg stop gap due to the periodic modulation of the effective index in the propagation direction [24].

In order to tune the structures, sequential conformal deposition of $HfO_2$ atomic layers was performed. $HfO_2$ was chosen as the ALD material due to its wide band gap, low optical absorption [25], and direct CMOS-compatibility having been used as a high-*k* dielectric gate insulator in the 45 nm technology node [26]. Prior to each deposition step, samples were cleaned with acetone, isopropanol and UV ozone. The UV generated ozone was used to create a hydrophilic surface favorable for the ALD process. Monolayer films were deposited at a temperature of 200°C using two precursors, tetrakis(diethylamido)hafnium(IV) [Hf(DEA)4] and water ($H_2O$) vapor, in alternating pulses. Nitrogen gas was flowed through the reaction chamber during the entire process. Lower temperature depositions are also possible with the trade-off of longer deposition times. The process is self-limiting and deposits one atomic layer at a time, with deposition rate of approximately 0.1 nm per minute. As shown by the SEM image in Figure 1(c), the ALD deposited film is high quality and uniform even inside of the air holes. Because ALD is a conformal process, each cycle incrementally decreases the hole radii and increases the slab thickness. The increase in brightness around the hole after ALD deposition is due to charging effects in the $HfO_2$ during SEM imaging.

Digital tuning was performed in increments of 40 atomic layers, with $1.05 \pm 0.05$ Å thickness for each $HfO_2$ atomic layer [27]. After each of these deposition steps (40 atomic layers), transmission measurements were performed for both the TE and TM polarizations. Light from a supercontinuum source was coupled into the on-chip polymer couplers using a polarization-maintaining tapered-lensed fiber. The output from the chip was similarly coupled to a tapered-lensed fiber and measured with an optical spectrum analyzer in the spectral range of 1300 to 1600 nm. The measured transmission spectra were normalized by the transmission spectra through a reference strip waveguide.

Figure 3(a) shows a series of TE transmission spectra after sequential ALD deposition steps. A wide bandwidth transmission region extends across the lower wavelength range followed by a sudden drop in transmission around 1514 nm for the pre-deposition measurement. The slow-light regime, which is close to onset of the waveguiding mode, is characterized by high group-indices as shown in figure 4(a). We observed a deterministic red-shift in the slow-light TE-like mode onset edge from 1513.8 nm (before ALD tuning) to 1533.7 nm (after 160 ALD deposition cycles), with the slow-light edge determined by a 10-dB drop in transmission corresponding to a group index of approximately 40. The inset of Figure 4(a) illustrates that the red-shift is linear, with a 140 ± 10 pm per monolayer control of the slow-light mode onset edge. The initial deposition step was not used in calculating this value because the slow-light red-shift in the first deposition step was slightly smaller than subsequent deposition steps. This is likely due to the formation of an 8-10 Å interfacial layer between $HfO_2$ and silicon during the first 20 ALD deposition cycles [27, 28]. In addition, on a different chip we have also tuned the slow-light edge across the entire optical communications C-band (and into part of the L-band), with tuning from 1530.6 nm to 1597.8 nm with 450 ALD cycles, i.e. 150 ± 10 pm per monolayer.

Figure 3(b) shows a series of TM transmission spectra after sequential ALD deposition steps. Unlike the TE-like slow-light mode which is found in the TE band gap, the TM-like slow-light modes are found on either side of the stop gap (illustrated by the dashed-lines in the computed band structure of Figure 2(b)). The TM-like modes are also red-shifted with the ALD deposition. The red-shift for the shorter-wavelength TM slow-light mode is likewise linear with control from 1370.7 nm to 1403.5 with 160 ALD layers, or 250 ± 10 pm per monolayer. The larger TM shift is due to the larger modal area and overlap with the $HfO_2$ monolayers. In comparing the slow-light tuning of the TE and TM modes, we note that there is a differential shift of 110 ± 30 pm per monolayer. This difference can be used for exact tuning of the pump-Stokes frequency spacing, in order to match the optical phonons (15.6-THz) in single-crystal silicon, for cross-polarized Raman amplification [24].

Along with PhC waveguide transmission measurements, MZI transmission measurements were taken to determine group indices using a frequency-domain interferometric technique [3,21]. The MZI structure is shown in Figure 1(b). The results of the measurements (over a total of 160 atomic layers) are summarized in Figure 4(a). The solid lines are from exponential fitting of group indices which were in-turn deduced from the spectral positions of minima and maxima in the MZI transmission with: $n_g(\lambda) = \lambda_{min} \lambda_{max} / (2L(\lambda_{min} - \lambda_{max})) + n_g^{ref}$, where $L$ is the PhC waveguide length [3]. Both before and after ALD controlled tuning, group indices of more than 60 were consistently obtained in our measurements.

Furthermore, higher order dispersion was also studied because of the important role it plays in pulse propagation in slow-light PhC waveguides [3,29]. At low group-velocities there can be a significant increase in temporal pulse width and pulse shape asymmetry due to higher order dispersion [30]. We investigated the effects of ALD on higher order dispersion, particularly the group velocity dispersion (GVD; $\lambda^2/2\pi c)(\partial n_g/\partial \lambda)$ and third-order dispersion (TOD; $\lambda^2/2\pi c)(\partial (GVD)/\partial \lambda)$. The GVD and TOD results for the different ALD deposition steps are summarized in Figures 4(b) and 4(c) respectively. The results show that the GVD and TOD do not change appreciably while the slow-light mode onset edge is deterministically tuned by ALD, with a variation of only 3% when determined from experimental group-index data which has been fitted. This small variation is not necessarily due to ALD tuning but can also originate in part from uncertainty in fitting the data.

Surface roughness is recognized as a significant factor in PhC waveguide losses [31]. Previous work [32] using similar waveguides has demonstrated a decrease in propagation loss as the size of the hole is decreased, which is attributed to the smaller surface area which results in reduced scattering. Ultra-smooth ALD films have also been demonstrated to not significantly affect the quality factors of optical resonators [19]. Atomic force microscopy studies by Puthenkovilakam et al. [28] have confirmed deposited $HfO_2$ films can have a root-mean-square roughness as small as 2.7Å, which is much smaller than the silicon sidewall roughness of the PhC structures as fabricated [33]. If extra surface roughness was introduced due to the deposition, an increase in propagation losses would steadily reduce the maximal group indices which can be resolved by the interferometric measurements. Since the ranges of group-index values extracted are comparable for each deposition thickness, our measurements suggest that the addition of ALD HfO$_2$ to the PhC waveguide structure does not introduce extra roughness or increase the propagation losses.

Another factor which can affect the propagation loss of PhC waveguides is the spatial confinement of slow-light modes. Increased field overlap with the sidewall will cause increased light scattering in the presence of roughness [31]. The effects of ALD on confinement were studied for various group indices using 3D plane wave expansion computations [23]. Figure 5 shows the average effective modal area of slow-light modes for two different group-indices (of 50 and 6) at different deposition steps. Figure 5(b) shows a cross-section of the slow-light mode (at a group-index of 50) through the air-clad silicon PhC waveguide before deposition and after deposition of 160 HfO$_2$ atomic layers. Note that the slow-light mode is less localized than the waveguide mode at a group-index of 6, suggesting potentially higher sensitivity to ALD deposition. For the group-index of 50, the effective modal area increases from $1.79 \times 10^{-13}$ m$^2$ to $2.03 \times 10^{-13}$ m$^2$, i.e. by 13.4% after 160 atomic layers. For a group-index of 6, the effective modal area also increases with ALD depositions (from $1.24 \times 10^{-13}$ m$^2$ to $1.32 \times 10^{-13}$ m$^2$), albeit a smaller increase of 6.4% compared to the slow-light mode. As discussed previously, the deposition is conformal and does not introduce extra roughness; thus, the fabrication-induced roughness stays the same (or decreases) with increasing number of layers. However, the higher refractive index of HfO$_2$ compared to air results in a slightly graded structure wherein the mode interacts less with the surface. As shown in Figures 5(a), the average effective modal area increases at a slower rate than the increase in cross-sectional area of the dielectric material due to the conformal ALD coating (25.5% after 160 ALD layers)

Thus, the conformal deposition reduces Rayleigh scattering thereby reducing the propagation loss.

In conclusion, we have demonstrated the control of slow-light dispersion characteristics of W0.9 PhC waveguides using a self-limiting monolayer precision process. High group indices were digitally tuned with sequential atomic layer depositions without increasing propagation losses. A red-shift of 140 ± 10 pm per atomic layer was observed for the slow-light mode onset edge, while no appreciable change was observed in the GVD and TOD. A differential shift of 110 ± 30 pm per monolayer in slow-light tuning of the TE and TM modes was observed. This difference can be used for exact tuning of the pump-Stokes frequency spacing for Raman amplification. As a low temperature post-fabrication process, the atomic layer deposition of $HfO_2$ is an enabling passive tuning technology for many practical chip-scale slow-light devices and modules.

This work was partially supported by DARPA, the New York State Foundation for Science, Technology and Innovation, and the National Science Foundation (ECCS-0622069 and CHE-0641523). The authors kindly thank J. F. McMillan and M. Eizenberg for useful discussions.

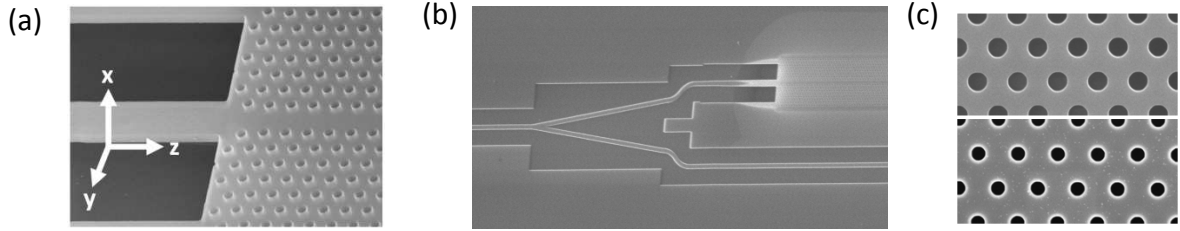

FIG. 1. (color online). SEM image of (a) PhC waveguide and strip waveguide interface (b) MZI structure with PhC waveguide on upper section and strip waveguide on lower section. (c) PhC before deposition (upper image) and after 160 ALD layers of $HfO_2$.

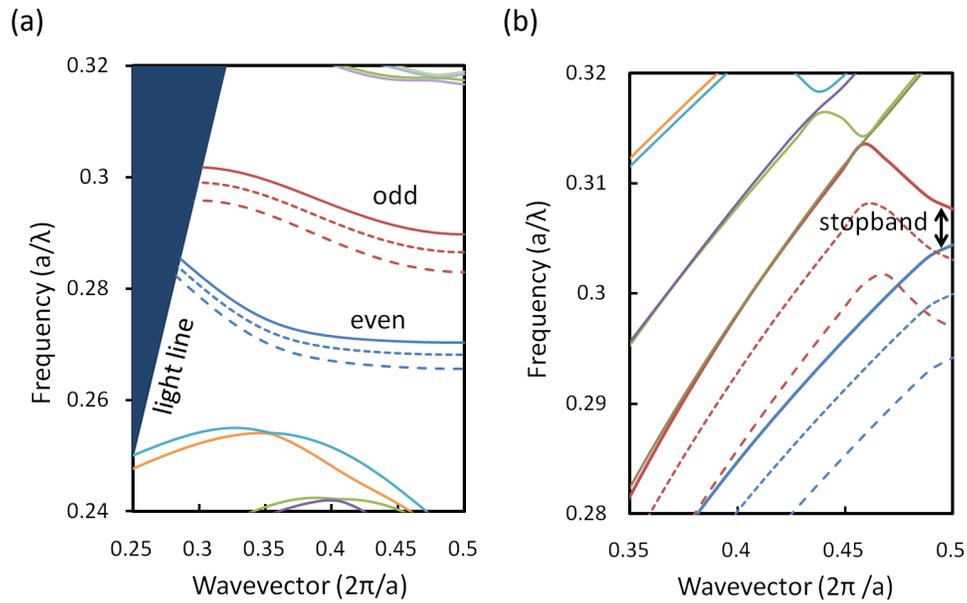

FIG. 2. (color online). Band structures calculated using plane-wave expansion method. Solid-lines correspond to no $HfO_2$ deposition. Dotted-lines and dashed-lines correspond to 80 and 160 atomic layers of $HfO_2$ deposition, respectively. (a) Projected band diagram for TE-like W0.9 waveguide modes (b) Corresponding TM band diagram. Note that the vertical and horizontal ranges are different than in (a).

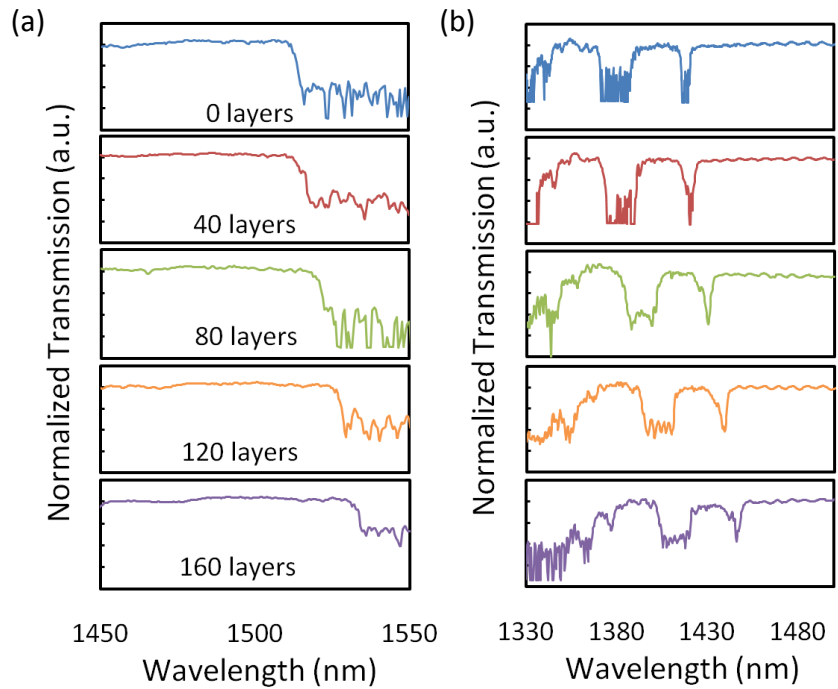

FIG. 3. (color online). (a) TE transmission measurements for different ALD tuning steps. The different colors denote the various ALD layers. (b) Corresponding TM transmission measurements [same color code as in (a)].

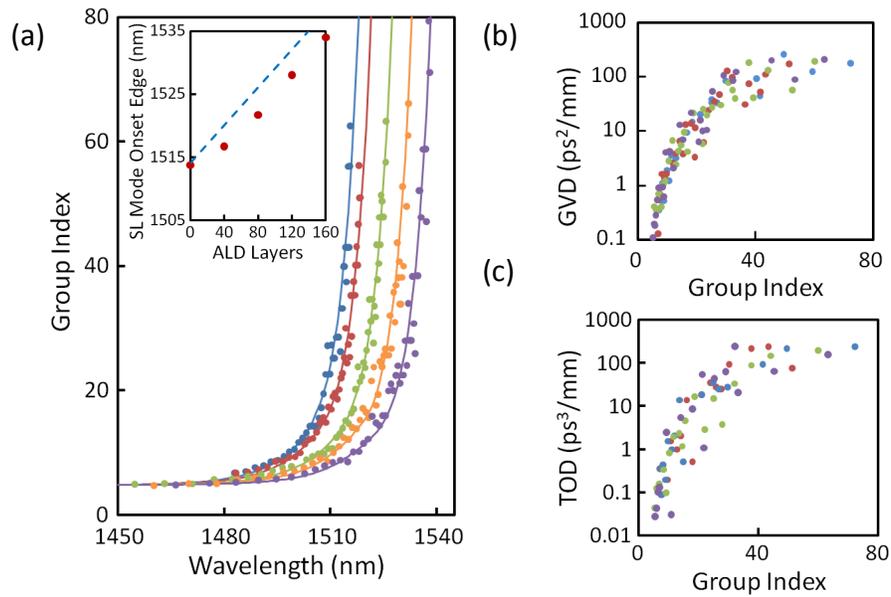

FIG. 4. (color online). (a) Group-index measurements from MZI devices. The number of atomic layers deposited is indicated by the color of the circles. Refer to the color code described in Figure 3(a). The solid lines are provided for clarity. Inset: measured slow-light (SL) mode onset (red circles) and numerical simulations (blue dashed line). (b and c) Measured group-velocity dispersion and third-order dispersion for the different ALD tuning steps.

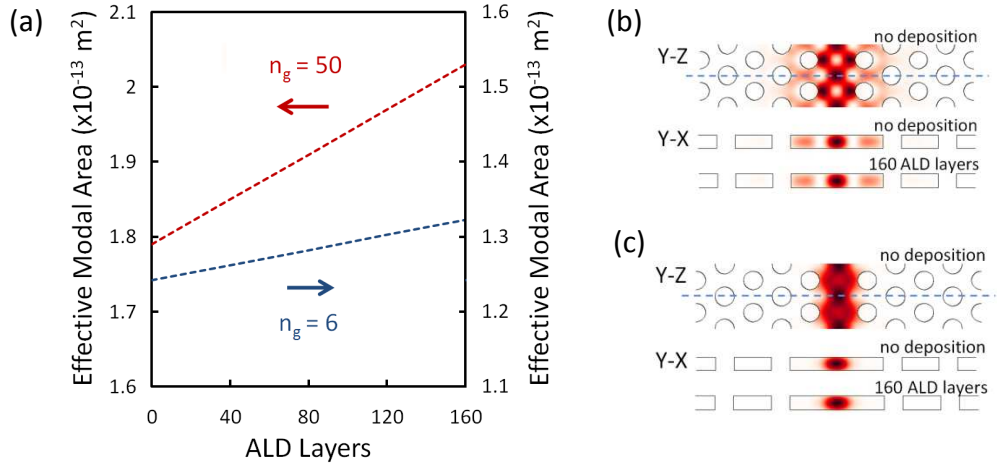

FIG. 5. (color online). (a) Average effective area of PhC waveguide slow light modes with ALD deposition at group-index of 50 (red line) and group-index of 6 (blue line). Calculated modes of W0.9 waveguide at (b) group-index of 50 (c) group-index of 6. The blue dotted line indicates where the cross-sections are taken perpendicular to the Y-Z plane. The resulting Y-X cross-sections before and after 160 atomic layers of $HfO_2$ have been deposited are shown below the Y-Z figures.